\documentclass[useAMS,usenatbib]{mn2e}

\usepackage{txfonts} 
\usepackage{epsfig} 
\usepackage{graphicx}
\usepackage{natbib} 

\newcommand{\kms}{{\km\second^{-1}}}

\newcommand{\km}{\,{\textnormal{km}}}
\newcommand{\second}{\,{\textnormal{s}}}

\newcommand{\dechms}[4]{$#1^{\rm h}#2^{\rm m}#3\mbox{$^{\rm s}\mskip-7.6mu.\,$}#4$}

\newcommand{\decdms}[4]{$-#1^{\circ}#2'#3\mbox{$''\mskip-7.6mu.\,$}#4$}

\newcommand{\Td}    {T_\mathrm{d}}

\newcommand{\mo}    {$M_{\sun}$}

\newcommand{\methanol}  {CH$_3$OH}

\newcommand{\ethanol}  {CH$_3$CH$_2$OH}

\newcommand{\acetaldehyde} {CH$_3$CHO}
\newcommand{\methylformate} {CH$_3$OCHO}

\newcommand{\eg}    {e.\,g.,}

\newcommand{\apj}{ApJ}
\newcommand{\araa}{ARAA}
\newcommand{\aap}{A\&A}
\newcommand{\apjl}{ApJ}
\newcommand{\aj}{AJ}
\newcommand{\mnras}{MNRAS}

\begin{document}

\title{ALMA reveals a candidate hot and compact disk around the O-type protostar IRAS 16547$-$4247}

\author[Zapata et al.]{Luis A.\ Zapata$^1$,  Aina Palau$^1$, Roberto Galv\'an-Madrid$^{1,2}$, Luis F.\ Rodr\'{\i}guez$^1$, 
\newauthor Guido Garay$^{3}$, James M. Moran$^{4}$, and Ramiro Franco-Hern\'andez$^{3}$ \\ 
$^{1}${Centro de Radiostronom\'{\i}a y Astrof\'{\i}sica, Universidad Nacional Aut\'onoma de M\'exico, 58089 Morelia, Michoac\'an, M\'exico}\\
$^{2}${European Southern Observatory, Karl-Schwarzschild-Str. 2, D-85748 Garching, Germany}\\
$^{3}${Departamento de Astronom\'\i a, Universidad de Chile, Casilla 36-D, Santiago, Chile}\\
$^{4}${Harvard-Smithsonian Center for Astrophysics, 60 Garden Street, Cambridge, MA 02138}}

\date{Accepted \today. Received \today; in original form \today}

\pagerange{\pageref{firstpage}--\pageref{lastpage}} \pubyear{2014}

\maketitle

\label{firstpage}


\begin{abstract}
We present high angular resolution ($\sim$ 0.3$''$) submillimeter continuum (0.85 mm) and line observations of the  
O-type  protostar  IRAS 16547$-$4247 carried out with the {\it Atacama Large Millimeter/Submillimeter Array} (ALMA). 
In the 0.85 mm continuum band, the observations revealed two compact sources (with a separation of 2$''$), 
one of them associated with IRAS 16547$-$4247, and the other one to the west. Both sources are well resolved angularly, revealing a clumpy structure.  
On the other hand, the line observations revealed a rich variety of molecular species related to both continuum sources. In particular,
we found a large number of S-bearing molecules, such as the rare molecule methyl mercaptan (CH$_3$SH).
At scales larger than 10,000 AU, molecules ({\it e.g.}, SO$_2$ or OCS) 
mostly with low excitation temperatures in the upper states (E$_k$ $\lesssim$ 300 K) are present in both millimeter continuum sources, and 
show a southeast-northwest velocity gradient of 7 $\kms$ over 3$''$ (165 $\kms$ pc$^{-1}$). We suggest that this gradient probably is produced by the 
thermal (free-free) jet emerging from this object with a similar orientation at the base. At much smaller scales (about 1000 AU),
molecules with high excitation temperatures (E$_k$ $\gtrsim$ 500 K) are tracing a rotating structure elongated perpendicular to the orientation 
of the thermal jet, which we interpret as a candidate disk surrounding IRAS 16547$-$4247. The dynamical mass corresponding to the velocity gradient of the candidate to disk is about 20 \mo, which is consistent with the bolometric luminosity of IRAS 16547$-$4247.
\end{abstract}


\section{Introduction}

Recent theoretical studies by \citet{kru2009}, \citet{pet2010}, and \citet{kui2010} have demonstrated that stars up to 140 M$_\odot$ 
can be formed in a similar way to the low- and intermediate-mass stars, that is, through flattened accretion disks. 
The existence of disks around high-mass stars solves the long-standing radiation pressure problem, where 
the powerful radiation pressure of the star is expected to halt the infalling material.  However, at this point, there is a lack of detections of
centrifugally supported circumstellar disks surrounding young O-type stars due probably to the large distances of
the stars (a few kpc), their very small number, and the complexity of the regions of massive star formation \citep[\eg][]{zin2007, zap2011, nar2012}.   

There are a few cases where disks rotationally supported are present in young O-type stars associated with a large luminosity 
(of about 10$^{4-5}$ L$_\odot$):  AFGL2591-VLA3, \citet{wan2012,jim2012}; IRAS 18360-0537, \citet{qiu2012};  
NGC 7538 IRS1, \citet{mos2014,kla2009,hof2012}; W51 North, \citet{zap2009,zap2010}; W33A, \citet{gal2010}.   
In the cases of  IRAS 18360-0537, NGC 7538 IRS1, and W51 North there is some evidence that the innermost parts of the 
disks are Keplerian, similar to the rotating structures found in low-mass stars.  
It is important to note that most of these structures are $\gtrsim$ 5000 AU in size (with the exception of AFGL 2591 and NGC 7538), 
and most likely are forming systems of massive stars .

\begin{figure*}
\begin{center}
\includegraphics[scale=0.40]{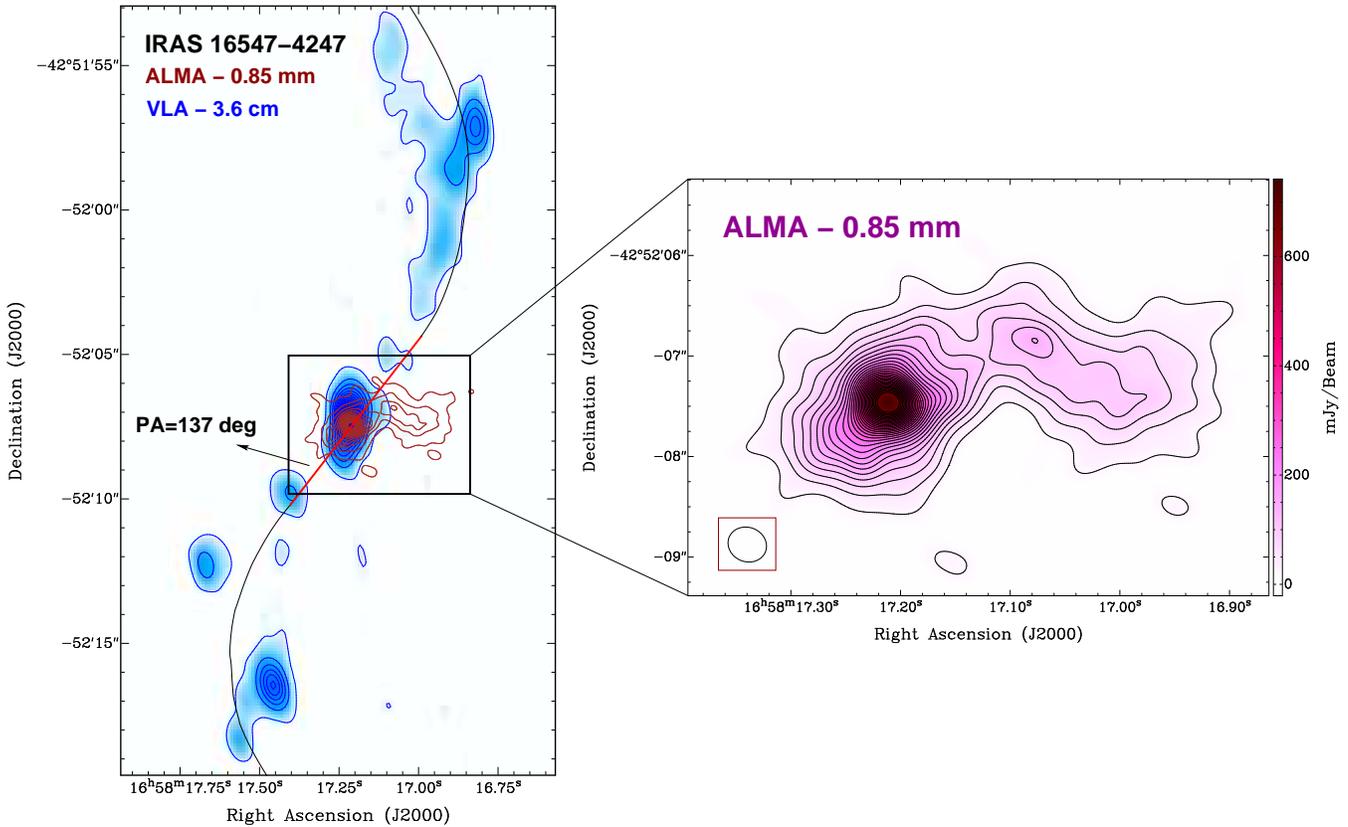}
\caption{ 3.6 cm and 0.85 mm continuum color-scale/contours images from IRAS 16547 obtained with 
the {Very Large Array} \citep{rod2008} and {The Atacama Large Millimeter/Submillimeter Array} 
(this work), respectively.  The blue color-scale/contours show the continuum emission 
arising at 3.6 cm, while the violet color-scale/contours show the continuum emission at 0.85 mm. Note that the 3.6 cm 
continuum emission traces a radio thermal (free-free) jet with an orientation northwest-southeast 
\citep{rod2008, gar2003}, while the emission detected by ALMA is only present in the middle and traces 
dust thermal emission.  The blue contours range from 5\% to 90\% of the peak emission, in steps of 5\%. 
The peak of the centimeter continuum emission is 6 mJy beam$^{-1}$.  The brown contours range 
from 5\% to 90\% of the peak emission, in steps of 5\%. The violet contours (on the right panel) range 
from 6\% to 90\% of the peak emission, in steps of 3\%. 
The peak of the millimeter continuum emission is 
0.67 Jy beam$^{-1}$. The synthesized beam of the ALMA continuum image is shown in the lower left corner, 
in the right image. The color-scale bar on the right panel indicates the peak flux in mJy beam$^{-1}$. 
The P.A. of the radio thermal jet close to the
source is estimated to be around 137$^\circ$, and the antisymmetric lines shows the jet trajectory associated 
with a precessing source model \citep{rod2008}. The r.m.s.\ noise for the ALMA continuum image is 7 mJy beam$^{-1}$.}
\label{fig1}
\end{center}
\end{figure*}

IRAS 16547$-$4247 ({\it hereafter IRAS 16457}) is catalogued as a young massive protostar with a bolometric luminosity of 
6.2 $\times$ 10$^4$ L$_\odot$ \citep{gar2003}, 
equivalent to that of a single O8 zero-age main-sequence star \citep{pan1973}, although it is probably a cluster for which the most massive source 
would have a slightly lower luminosity. The source is located at a distance of 2.9 $\pm$ 0.6 kpc \citep{rod2008}.  Observations at radio wavelengths 
have revealed a triple continuum source that is aligned in a northwest-southeast direction (position angle: P.A. $\sim$140$^\circ$), 
with the outer lobes symmetrically separated from the central source by an angular distance of $\sim$10$''$, equivalent to a physical 
separation in the plane of the sky of $\sim$0.14 pc \citep{gar2003}. The fit to the radio jet by  \citet{rod2008}  gives a position
angle of 137$^\circ$ within a few arc seconds from the star. It should be noted, however, that at smaller scales the core of the jet shows a 
somewhat different position angle of 164$^\circ$ \citep{rod2008}. In this paper, we will adopt as the representative position angle of the jet the 
value of 137$^\circ$.

The central source --related to the infrared source IRAS 16547-- has a positive spectral index which is consistent with that expected 
for a radio thermal (free--free) jet, while the spectral index of the lobes suggests a mix of thermal and non-thermal emission \citep{ara2007}. 
The APEX molecular observations of \citet{gar2007} revealed the presence of a massive and energetic bipolar outflow 
(flow mass $\sim$110 M$_\odot$; mass outflow rate $\sim$10$^{-2}$ M$_\odot$ yr$^{-1}$; momentum $\sim$10$^3$ \mo~km s$^{-1}$ 
and kinetic energy 10$^{48}$ erg)  with lobes $\sim$0.7 pc in extent and aligned with the thermal jet located 
at the center of the core.

Very Large Array (VLA) and Submillimeter Array (SMA) observations with high spatial resolution have revealed  
a rotating structure associated with IRAS 16547, traced at small scales ($\sim$50 AU) by the H$_2$O masers and 
at moderate scales ($\sim$1000 AU) by the thermal emission of SO$_2$. However, both rotating
structures have slightly different position angles and most likely are tracing different parts of IRAS 16547. The SO$_2$ 
rotating structure has an east-west orientation (P.A. $\sim$90$^\circ$), while the H$_2$O structure a P.A. $\sim$40$^\circ$. 
The poorly resolved structure of the SO$_2$ can be modelled as a rotating ring or two separate objects, while
the H$_2$O rotating structure, on the other hand,  was suggested to be a compact Keplerian 
disk surrounding IRAS 16547 \citep{fra2009}. 

In this study, we present submillimeter line and continuum observations, made with the 
{Atacama Large Millimeter/Submillimeter Array} (ALMA) of the massive protostar  IRAS 16547. 
We report on the detection of a candidate to compact and hot molecular rotating disk with a size of about 1000 AU and with an orientation 
perpendicular to the radio jet emerging from this object. 
In \S 2 and 3 we discuss the observations, and results. In \S 4 we present the discussion and conclusions.

\begin{figure*}
\begin{center}
\includegraphics[scale=0.7, angle=-90]{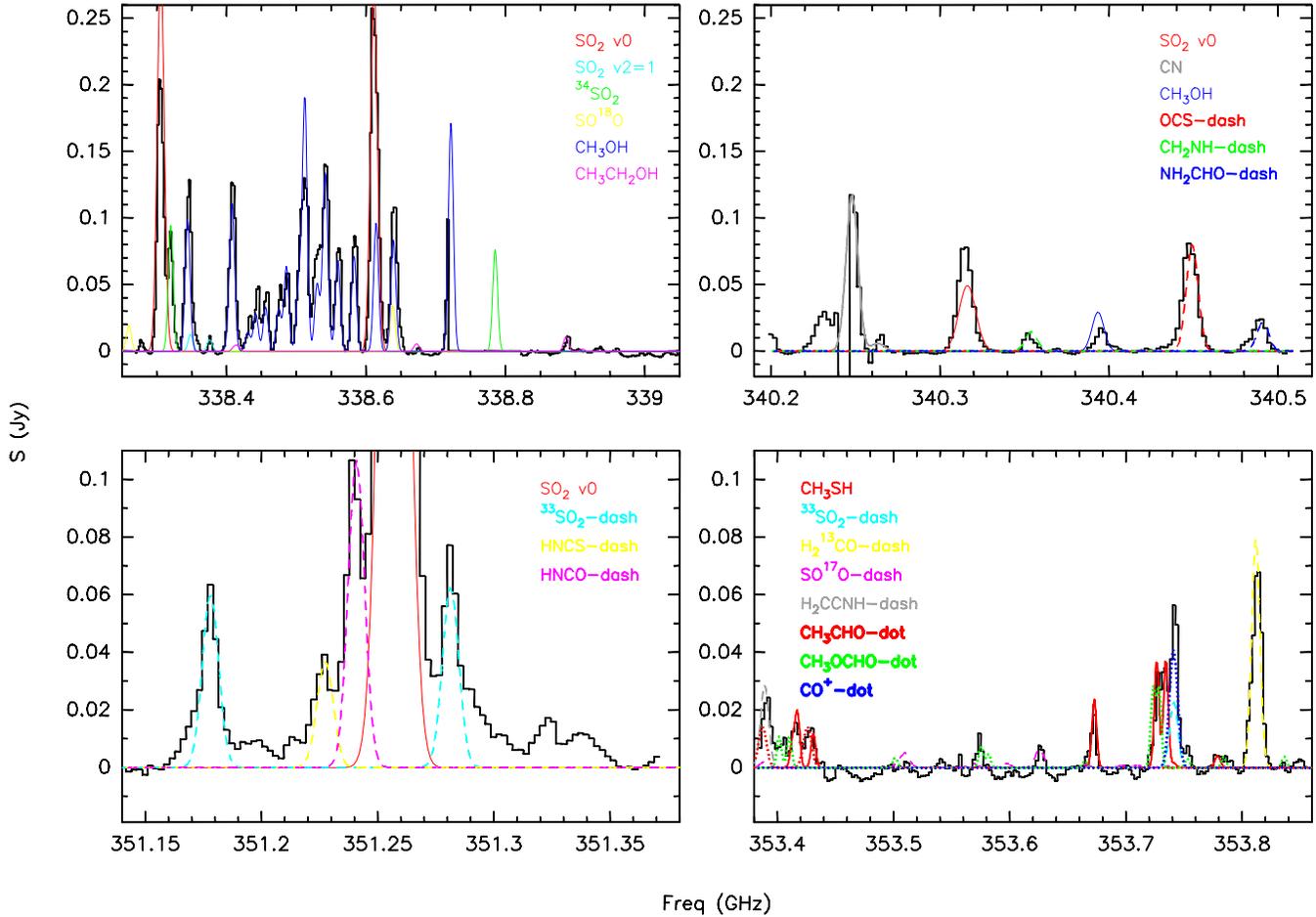}
\caption{ Average ALMA spectra obtained from IRAS 16547. The color lines represent the synthetic LTE spectra assuming 
the parameters described in the text. We remark that this synthetic spectra in only used for line identification.  }
\label{fig2}
\end{center}
\end{figure*}

\begin{figure}
\begin{center}
\includegraphics[scale=0.32]{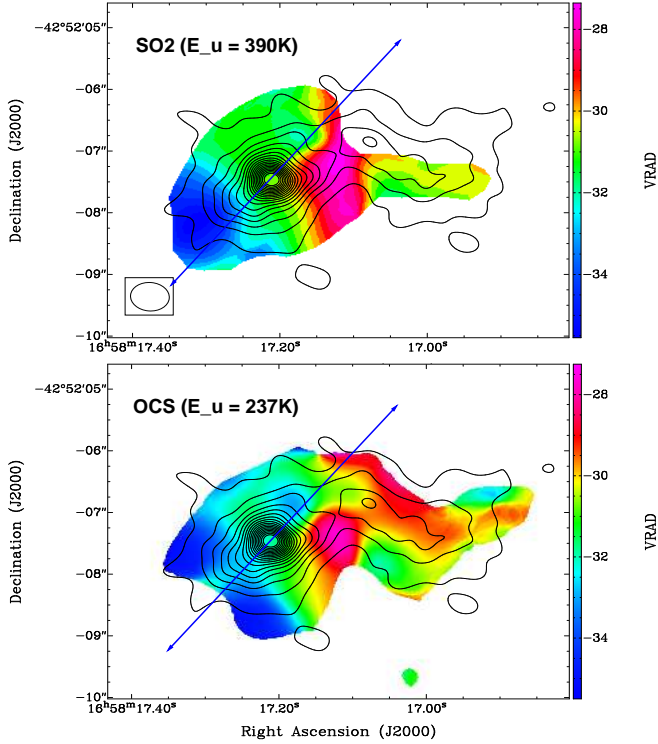}
\caption{ SO$_2$ v=0 28$_{2, 26}$ - 28$_{1, 27}$,	 
and OCS (28$-$27) thermal emission  integrated-intensity weighted 
velocity (moment one) color maps  overlaid with the 0.85 mm continuum maps from 
IRAS 16547. The black contours show the continuum emission arising at 0.85 mm.  
The contours range from 5\% to 90\% of the peak emission, in steps of 5\%. The peak of the 
millimeter continuum emission is 0.67 Jy beam$^{-1}$. The synthesized beam of the ALMA 
continuum image is shown in the lower left corner of the upper panel. The color-scale bar 
on the right indicates the LSR radial velocities in km s$^{-1}$. The LSR systemic velocity 
of IRAS 16547 is $-$30.6 $\kms$ \citep{gar2003}. The blue line in both panels 
marks the orientation of the thermal (free-free) jet \citep{rod2008, gar2003} with a P.A. 
estimated to be around 137$^\circ$.}
\label{fig3}
\end{center}
\end{figure}

\begin{figure}
\begin{center}
\includegraphics[scale=0.32]{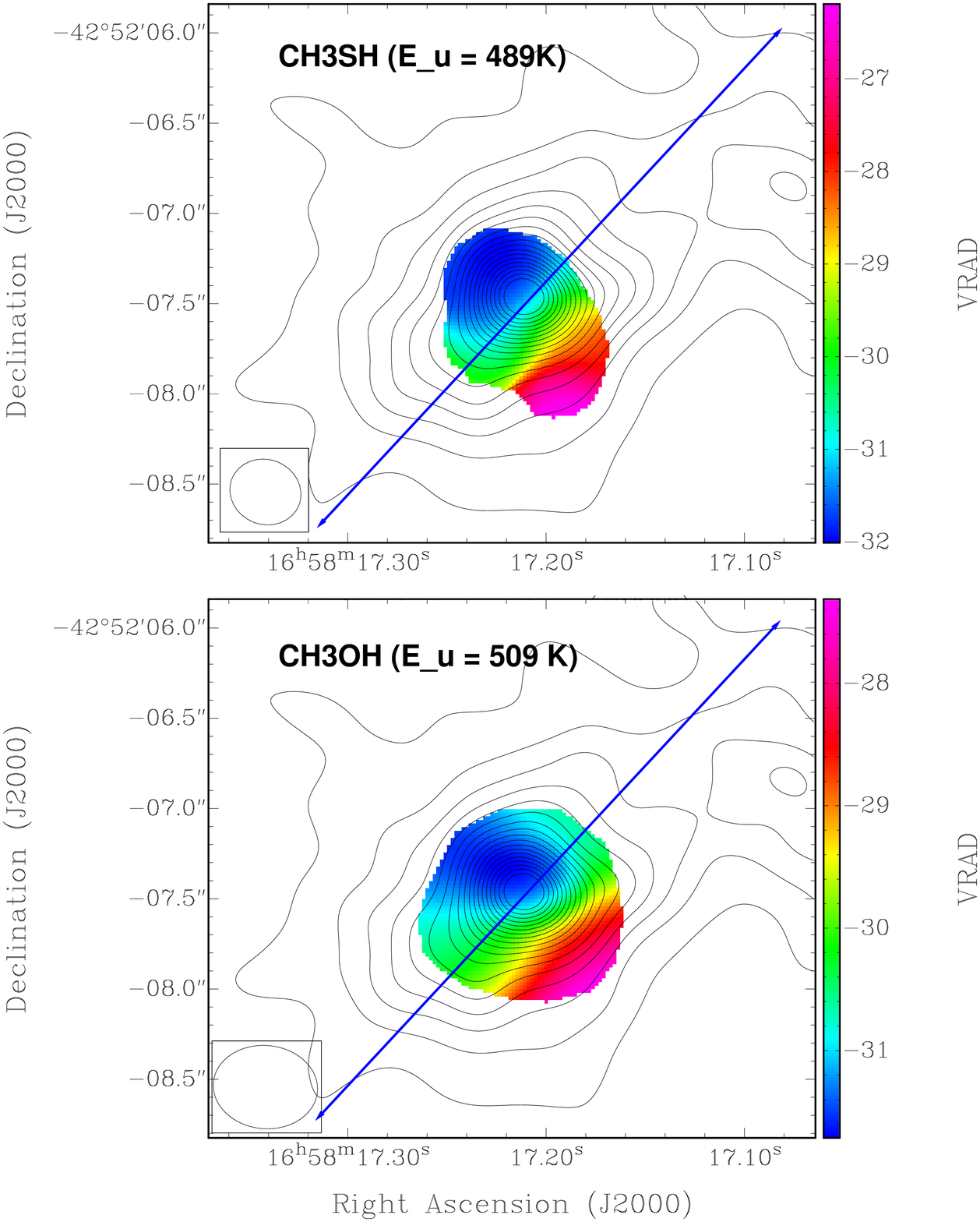}
\caption{ CH$_3$SH 14$_9$-13$_9$ $-$/+A and \methanol\ 16$_{6, 10}$ - 17$_{5, 13}$ $++$ integrated-intensity 
weighted velocity (first moments) color maps  overlaid with the 0.85 mm continuum maps from the 
innermost parts of IRAS 16547. The black contours show the continuum emission arising at 0.85 mm.  
The contours range from 5\% to 90\% of the peak emission, in steps of 5\%. The peak of the millimeter 
continuum emission is 0.67 Jy beam$^{-1}$. The synthesized beam of the ALMA continuum image is shown 
in the lower left corner. The color-scale bar on the right indicates the LSR radial velocities in km s$^{-1}$. 
The LSR systemic velocity of IRAS 16547 is $-$30.6 $\kms$ \citep{gar2003}. 
The blue line in both panels, marks the orientation of the thermal (free-free) jet \citep{rod2008, gar2003} 
with a P.A. estimated to be around 137$^\circ$. The border of the velocity map corresponds to 3-$\sigma$ level.}
\label{fig4}
\end{center}
\end{figure}   

\section{Observations}

The observations were carried out with 32 antennas of {The Atacama Large Millimeter/Submillimeter Array} 
(ALMA) on April (16 antennas) and August 2012 (21 antennas), during the cycle 0 science data program.  
The array at that point only included antennas with diameters of 12 meters.
The 496 independent baselines ranged in projected length from 25 to 364 m.  
  
The phase reference center for the observations was at $\alpha_{J2000.0}$ = 
\dechms{16}{58}{17}{24}, $\delta_{J2000.0}$ = \decdms{42}{52}{08}{09}, the position of the object IRAS 16547.
The primary beam of ALMA at 345 GHz has a FWHM $\sim17.6''$. The dusty emission from IRAS 16547 
falls well inside of the FWHM. The ALMA digital correlator was configured in 4 spectral windows of 1875 MHz and
3840 channels each.  This provides a spectral resolution of 0.488 MHz ($\sim$ 0.4 km s$^{-1}$) 
per channel. The spectral windows were centered at 350.355 GHz, 352.228 GHz, 340.117 GHz, and 341.971 GHz,
in order to detect different spectral lines as for example, SO, $^{34}$SO, and CS.  These lines had already been
reported toward this massive young stellar object using the single dish telescopes {\it SEST} and {\it APEX} \citep{gar2003,gar2007}.  

Observations of Titan provided the absolute scale for the flux density calibration. 
For the time-dependent gain calibration, the nearby quasar J0607$-$085 was observed 
approximately every 15 minutes. The quasar 3C279 was used for the bandpass calibration.
      
The data were calibrated, imaged, and analyzed using the Common Astronomy Software Applications \citep[CASA;][]{mc2007}.  
To analyze the data, we also used the KARMA \citep{goo96} and AIPS of the {National Radio Astronomy Observatory} (NRAO) software.  
In order to construct the continuum image, we used the total bandwidth of the observations (7.5 GHz) selecting only the line-free channels. However,
as there are many lines detected in the spectral windows, we have probably some contamination from very faint lines.  
 We used uniform weighting in the continuum and lines maps presented in this study.
The resulting r.m.s.\ noise for the continuum image was 7 mJy beam$^{-1}$ at an angular resolution of $0\rlap.{''}39$ 
$\times$ $0\rlap.{''}34$ with a P.A. = $69.2^\circ$.  For the line emission, the resulting r.m.s.\ noise was about 35 mJy beam$^{-1}$ $\kms$
at the same angular resolution.   
   
\begin{table*} 
\scriptsize
\centering
\caption{Transitions detected toward IRAS\,16547}
\begin{tabular}{l l c c r}
\\
\hline\hline\noalign{\smallskip}	
                                &         	&Rest Freq.     &$S_{ij}$\footnote{Line strength as given by the {\it Splatalogue}.}$\mu^2$ &E$_u$        \\
Molecule                 	&Transition   	&(GHz)        	&(D$^2$)                  							& (K)            \\
\\
\hline
\\
\noalign{\smallskip}

SO$_2$ v=0    		&18$_{4, 14}$--18$_{3, 15}$ 			&338.30599    	&26.81	&197\\
           			&20$_{1, 19}$--19$_{2, 18}$       		&338.61181 	&26.02        &199\\
           			&28$_{2, 26}$--28$_{1, 27}$          		&340.31641 	&32.05        &392\\
           			&5$_{3, 3}$--4$_{2, 2}$            		&351.25722 	&7.32        &36\\
SO$_2$ v$_2$=1   	&4$_{3, 1}$--3$_{2, 2}$ 				&338.34874    	&7.07        &792\\
           			&8$_{2, 6}$--7$_{1, 7}$            		&338.37638 	&5.09        &803\\
$^{34}$SO$_2$ v=0	&13$_{2, 12}$--12$_{1, 11}$ 			&338.32036    	&13.59        &93\\
$^{33}$SO$_2$ v=0	&8$_{4, 4}$--8$_{3, 5}$, F=19/2--19/2 	&351.17796    	&11.16        &73\\
           			&9$_{4, 6}$--9$_{3, 7}$, F=21/2--21/2 	&351.28137    	&12.88        &81\\
           			&19$_{4, 16}$--19$_{3, 17}$, F=41/2--41/2 	&353.74156    	&25.37        &217\\
SO$^{18}$O    		&20$_{0, 20}$--19$_{1, 19}$ 			&338.63882    	&43.41        &184\\
OCS            		&28--27 							&340.44927      &14.32        &237\\
HNCS $a$-type  	&30$_{1, 30}$--29$_{1, 29}$     		&351.22743      &80.60        &324\\
SO$^{17}$O    		&20$_{1, 20}$--19$_{0, 19}$ 			&353.62533    	&263.56        &186\\

CH$_3$SH      		&14$_{13}$--13$_{13}$ $-$/+A               	&353.41736  	&3.32 	& 864\\
				&14$_{13}$--13$_{13}$ E                  	&353.43107  	&3.33 	& 863\\
				&14$_9$--13$_9$ $-$/+A               		&353.67282  	&14.10 	& 479\\
				&14$_2$--13$_2$ $-$A               		&353.72626  	&23.60 	& 147\\
				&14$_8$--13$_8$ $-$/+A               		&353.73446  	&16.20 	& 406\\
				&14$_8$--13$_8$ E               			&353.74153  	&16.27 	& 408\\
				&14$_7$--13$_7$ E           	    		&353.77778  	&18.11  & 341\\
				&14$_{-7}$--13$_{-7}$ E           		&353.78050  	&18.11 	& 342\\
\hline				
\methanol\ 		&7$_{-1, 7}$--6$_{-1, 6}$ 			&338.34463    	&5.54   &71  \\
           			&7$_{0, 7}$--6$_{0, 6}$ + +               	&338.40868     	&5.66   &65  \\
           			&7$_{-6, 1}$--6$_{-6, 0}$ 			&338.43093     	&1.50    &254  \\
           			&7$_{6, 1}$--6$_{6, 0}$ + +               	&338.44234     	&1.49        &259  \\
           			&7$_{-5, 2}$--6$_{5, 1}$                		&338.45650     	&2.76       &189  \\
           			&7$_{5, 3}$--6$_{5, 2}$ 				&338.47529     	&2.76      &201  \\
           			&7$_{5, 3}$--6$_{5, 2}$ + +               	&338.48634     	&2.77       &203  \\
           			&7$_{2, 6}$--6$_{2, 5}$ $-$ $-$ 		&338.51286     	&5.23      &103  \\
           			&7$_{4, 3}$--6$_{4, 2}$ 				&338.53025     	&3.82     &161  \\
           			&7$_{3, 5}$--6$_{3, 4}$ + + 			&338.54080     	&4.60       &115  \\
           			&7$_{-3, 5}$--6$_{-3, 4}$         		&338.55993 	&4.64      &128  \\
           			&7$_{3, 4}$--6$_{3, 3}$                  		&338.58319     	&4.62       &113  \\
           			&7$_{1, 6}$--6$_{1, 5}$                  		&338.61500     	&5.68        &86  \\
           			&7$_{2, 5}$--6$_{2, 4}$ + + 			&338.63994     	&5.23      &103  \\
           			&7$_{2, 5}$--6$_{2, 4}$                  		&338.72163     	&5.14        &87  \\
           			&7$_{-2, 6}$--6$_{-2, 5}$         		&338.72294 	&5.20        &91  \\
           			&16$_{6, 10}$--17$_{5, 13}$ + +         	&340.39367     	&3.57        &509  \\

\hline
\ethanol\			&15$_{7, 8}$--15$_{6, 9}$                	&338.88621  	&13.67	&162\\
\hline
CN v=0        		&N=3--2, J=7/2--5/2, F=9/2--7/2     		&340.24777  	&9.01   &33\\
				&N=3--2, J=7/2--5/2, F=5/2--5/2     		&340.26177 	&0.59  	&33\\
				&N=3--2, J=7/2--5/2, F=7/2--7/2     		&340.26495 	&0.58  	&33\\
\hline
CH$_2$NH             	&3$_{1, 3}$--2$_{0, 2}$, F=4--3      		&340.35431  	&6.07 	&26\\
\hline
NH$_2$CHO         	&16$_{3, 14}$--15$_{3, 13}$         		&340.49109 	&605.53 &166\\
\hline
HNCO v=0              	&16$_{4, 13}$--15$_{4, 12}$          		&351.24085  	&30.93 	&794\\
\hline
H$_2$$^{13}$CO  	&5$_{0, 5}$--4$_{0, 4}$             		&353.81187 	&27.17 	&51\\
\hline
H$_2$CCNH       	&11$_{2, 10}$--12$_{1, 12}$          		&353.38912  	&4.25 	&97\\
\hline
\acetaldehyde\ $^{b}$	&19$_{0, 19}$--18${0, 18}$ E        		&353.38728    	&239.52 &172    \\
					&19$_{0, 19}$--18$_{0, 18}$ A        		&353.42595    	&239.30 &172    \\
\hline
\methylformate\ v=0\footnote{Marginal/blended detections}&29$_{4, 25}$--28$_{4, 24}$ E	&353.40185  	&74.20 	&274   \\ 
\methylformate\ v=0$^{2}$ &29$_{4, 25}$--28$_{4, 24}$ A            				&353.41059   	&74.21  &274   \\ 
\methylformate\ v=1$^{2}$ &28$_{7, 21}$--27$_{7, 20}$ A            				&353.57558   	&69.94  &461   \\ 
\methylformate\ v=0$^{2}$ &32$_{2, 31}$--31$_{2, 30}$ E            				&353.72360   	&84.18 	&290   \\ 
\methylformate\ v=0$^{2}$ &32$_{2, 31}$--31$_{2, 30}$ A            				&353.72862   	&84.15 	&290   \\ 
\hline
CO$^+$ 			&J=3--2, F=5/2--3/2          	&353.74126  	&2.40    &34   \\ 
\\
\hline
\hline
\end{tabular}
\end{table*}

\begin{figure}
\begin{center}
\includegraphics[scale=0.18]{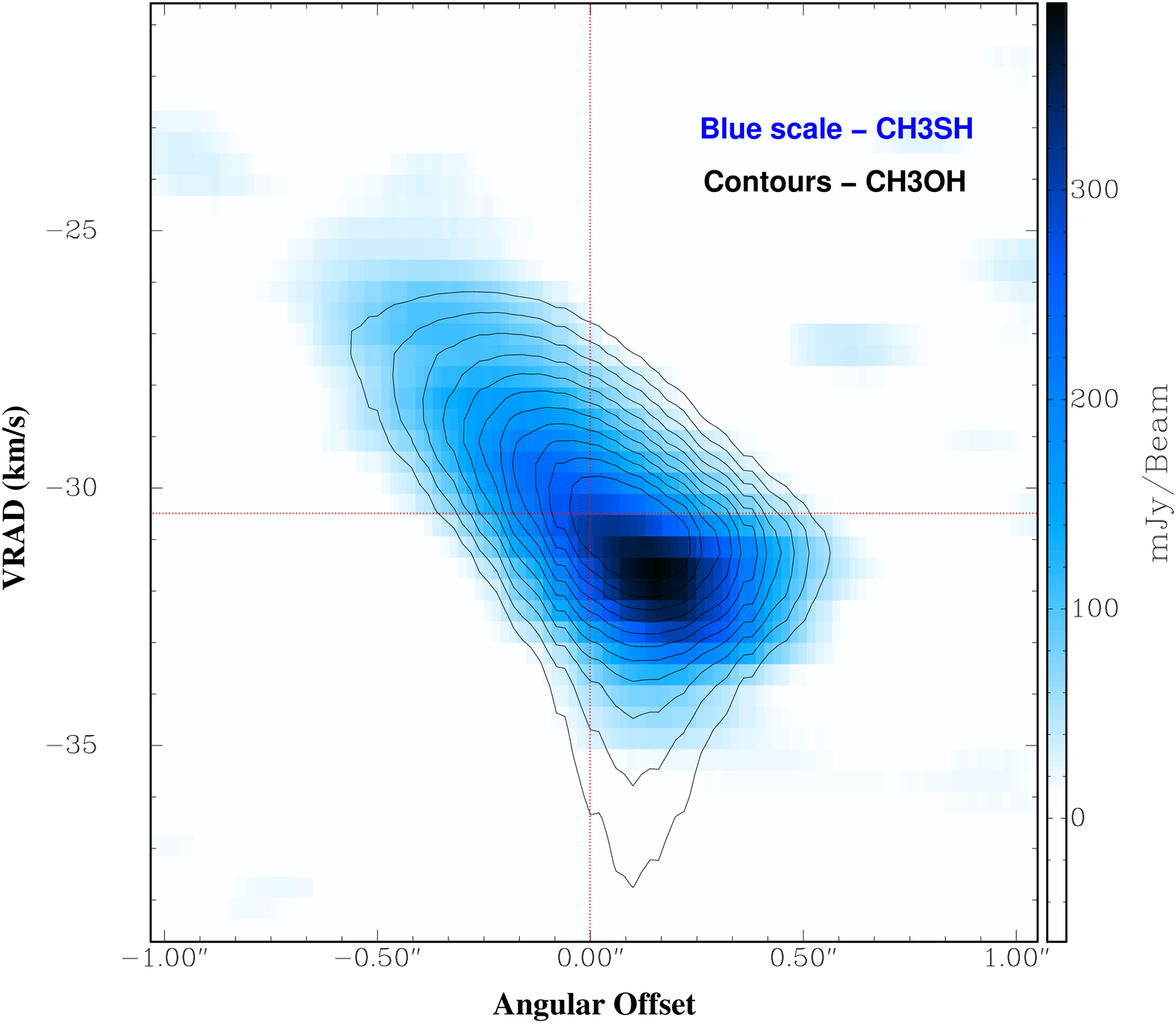}
\caption{ Position-velocity diagrams of the CH$_3$SH 14$_9$-13$_9$ $-$/+A (blue scale) 
 and \methanol\ 16$_{6, 10}$ - 17$_{5, 13}$ $++$
 (contours) emissions from the innermost parts of IRAS 16547. 
The contours range from 40\% to 90\% of the peak emission, in steps of 5\%. The peak of the 
 CH$_3$OH emission is 1.01 Jy beam$^{-1}$.   The color-scale bar on the right indicates 
the peak flux in mJy beam$^{-1}$. The LSR systemic velocity of IRAS 16547 is around 
$-$30.6 $\kms$ \citep{gar2003}.  The spectral and angular resolutions are described in the text. 
The P.A. for both PV-diagrams (CH$_3$SH and \methanol\ ) about 40$^\circ$. 
The units of the y-axis is in km s$^{-1}$. The angular offset here is relative to the peak of the continuum emission. 
The positive offsets are measured towards the NE. The 1-$\sigma$ RMS level corresponds to 35 mJy beam$^{-1}$. 
The radial velocity here is LSR.}
\label{fig5}
\end{center}
\end{figure}   
   
\section{Results}

\subsection{0.85 mm continuum emission}

In Figure 1, we present the main results of the ALMA 0.85 mm continuum observations of IRAS 16547.
The submillimeter source reported by  \citet{gar2003} was found to have a western extension by \citet{fra2009}. We resolved the source 
into two continuum sources with a clumpy morphology, one of them associated with the infrared source IRAS 16547, 
and the other one 2$''$  to its west (IRAS 16547-W). IRAS 16547-W was already reported by \citet{fra2009}
and we find that its peak lies $\sim$0.4$''$ to the west of the centimeter source IRAS 16547 D \citep{rod2008}.

The continuum source associated with IRAS 16547 is at a position of $\alpha_{J2000.0}$ = \dechms{16}{58}{17}{211}, 
$\delta_{J2000.0}$ = \decdms{42}{52}{07}{47}. For IRAS 16547-W, it is difficult to obtain a position because of its 
extended clumpy morphology. The central position of IRAS 16547-W is $\alpha_{J2000.0}$ = \dechms{16}{58}{17}{051}, 
$\delta_{J2000.0}$ = \decdms{42}{52}{07}{11}. Using a gaussian fitting we found for IRAS 16547 a 
flux density and peak intensity values of  2.1$\pm$0.10 Jy  and 675$\pm$30 mJy beam$^{-1}$, respectively.
The gaussian fitting was only made in the most central part of IRAS 16547, we do not include the SE-NW extended emission. 

For IRAS 16547-W, we found a flux density and peak intensity values of  1.6$\pm$0.20 Jy  and 140$\pm$30 mJy beam$^{-1}$, 
respectively. We also find from these fits
that the IRAS 16547 source has a deconvolved size of $0\rlap.{''}56$ $\pm$ $0\rlap.{''}02$ $\times$ $0\rlap.{''}50$ $\pm$ $0\rlap.{''}02$  with 
a P.A. = $+$115$^\circ$ $\pm$ 5$^\circ$. Therefore, at the distance of this object (2.9 kpc), the size of the continuum source 
is about 1620 AU $\times$ 1450 AU.

Assuming that the dust is optically thin and isothermal, and following \citet{hil1983}, the dust mass (M$_d$) will be 
directly proportional to the flux density (S$_\nu$) as:

\begin{equation}
M=\frac{S_\nu D^2}{B_\nu(\Td)\kappa_\nu},
\end{equation}
where $S_\nu$ is the flux density at the frequency $\nu$, $D$ is the distance to the Sun, $B_\nu(\Td)$ is the Planck function 
at the dust temperature $\Td$, and $\kappa_\nu$ is the absorption coefficient per unit of total (gas+dust) mass density. 
Writing Eq.~(1) in practical units \citep{pau2013} :

\begin{equation}
\left[\frac{M}{M_{\odot}}\right]=3.25\times
\frac{e^{0.048\,\nu/\Td}-1}{\nu^3 \kappa_\nu}\times
\left[\frac{S_\nu}{\mathrm{Jy}}\right]
\left[\frac{D}{\mathrm{pc}}\right]^{2},
\end{equation}

\noindent where $\Td$ is in K, $\nu$ is in GHz, and $\kappa_\nu$ is in cm$^2$g$^{-1}$. Taking a gas-to-dust ratio of 
100, a distance of 2.9 kpc, a dust temperature of 250 K \citep{her2014} for IRAS 16547 and 150 K 
(this value is uncertain and is obtained more or less from the excitation temperatures of some the lines present in this object) 
for IRAS 16547-W, and a dust mass opacity $\kappa_{850 \micron}$ = 0.015 cm$^2$ g$^{-1}$ (taking into account the gas-to-dust ratio) 
\citep{oss1994}. We estimate the total mass associated with the dust continuum emission related to the disk in IRAS 16547 of 
6.0 M$_\odot$ (integrating in an area $\sim$ 1$''$), and
for IRAS 16547-W a mass of 8.0 M$_\odot$. 
 The values of the masses obtained here for IRAS 16547 
and IRAS 16547-W have uncertainties of at least a factor two, due mainly to the error in the
determination to the dust mass opacity coefficient at this wavelength.

\subsection{Line thermal emission}
 
In Table 1 we show all spectral lines detected in these observations. 
We detected 15 molecular species with different transitions and their isotopologues.  
In particular, we found a large number of S-bearing molecules, for example the rare molecule methyl mercaptan (CH$_3$SH).
This molecule has been detected only in two star forming regions \citep[G327 and B1;][]{gib2000, cer2012}, other than the Galatic Center \citep{lin1979}
and Orion-KL \citep{kol2014}, and is supposed to trace hot regions were desorption from grain mantles is very efficient \citep{cer2012}. 

In order to properly identify the lines detected in the four ALMA spectral windows, we built synthetic spectra for all the possible 
molecules with transitions within $\pm3$~MHz of a given line, and compared the synthetic spectra with the observed 
spectra in the 4 observed windows, see Figure 2 and Table 1. This allowed us to take into account blending and contributions from transitions of 
different molecules to a given line. The synthetic spectra were computed assuming Local Thermodynamic Equilibrium 
and optically thin emission as in \citet{pal2011}, and using the molecular data from the Jet Propulsion Laboratory 
\citet{pic1998} and The Cologne Database for Molecular Spectroscopy \citet{mul2005}. Frequencies and energy 
levels of methyl mercaptan were taken from the Spectral Line Atlas of Interstellar Molecules 
(SLAIM, the catalog of Frank Lovas, accessible only through Splatalogue\footnote{http://www.splatalogue.net/}). To build the synthetic 
spectra, we adopted a linewidth (FWHM) of 7 km s$^{-1}$ (except for the SO$_2$ lines, for which we used a linewidth of 10 km s$^{-1}$), 
and used rotational temperatures in the range 70--300 K \citep{her2014}. We then varied the column 
density until the synthetic spectra reasonably matched the observed spectra.

In Figure 3 we show the first moment or the intensity weighted velocity of the  SO$_2$ v=0 28$_{2, 26}$ - 28$_{1, 27}$,	 
and OCS (28$-$27) thermal emission. Emission from these two molecules are present in IRAS 16547
as well as IRAS 16547-W. Both molecules emit at similar systemic LSR velocities.  The LSR systemic velocity 
of IRAS 16547 is $-$30.6 $\kms$ \citep{gar2003}.  In both molecules, the thermal emission 
associated with IRAS 16547 is stronger and more compact than that present in IRAS 16547-W.
IRAS 16547 shows a clear velocity gradient of 7 $\kms$ over a distance of 3$''$ (equivalent to 165 $\kms$ pc$^{-1}$).
\citet{fra2009} also reported the detection of the SO$_2$ at 1 mm in IRAS 16547 with a smaller velocity gradient 
(2 $\kms$ over 3$''$ or 47.6 $\kms$ pc$^{-1}$) at similar scales. This small
velocity gradient is probably caused by the lower sensitivity of the Submillimeter Array compared to ALMA.
Convolving the ALMA SO$_2$ map to the angular resolution reported in \citet{fra2009} revealed similar results.
We also noted that the SO$_2$ line mapped in SMA study, has an excitation energy much higher than that observed by \citet[][]{fra2009} (118 K).
The blueshifted gas emission is located to its southeast, while the redshifted gas velocities is at its northwest. 
In these new ALMA observations, the velocity gradient is seen in both molecules (SO$_2$ and OCS), 
with some redshifted gas associated with IRAS 16547-W. 

The first moment or the intensity weighted velocity of the CH$_3$SH 14$_9$-13$_9$ $-$/+A and \methanol\ 16$_{6, 10}$ - 17$_{5, 13}$ $++$
thermal emission is presented in Figure 4. The emission from these two molecules is exclusively present in IRAS 16547, and 
is resolved with dimensions of $0\rlap.{''}5$ $\times$ $0\rlap.{''}3$ with a P.A.=$40^\circ$, corresponding 
to 1450 AU  $\times$ 870 AU. There is a clear velocity gradient in both molecules and with an orientation 
almost perpendicular to that of the thermal jet and the 
molecules SO$_2$ and OCS. The blueshifted gas emission is found to its northeast, 
while the redshifted is at its southwest. 

The kinematics of the gas emission traced by the molecules CH$_3$SH 14$_9$-13$_9$ $-$/+A and \methanol\ 16$_{6, 10}$ - 17$_{5, 13}$ $++$ is
presented in the Figure 5. The position angle where this diagram was computed is 40$^\circ$. This reveals more clearly the velocity gradient 
observed in Figure 4 with a magnitude of 8 $\kms$ over 1$''$ (equivalent to 566 $\kms$ pc$^{-1}$). A small contribution 
of Keplerian velocities is observed to the northeast of the source. 
The reason why we do not see the full Keplerian motions in this position-velocity diagram is probably because we need to better 
resolve the innermost parts of the possible disk, where the velocities increment more rapidly. However, there is also the possibility that 
the motions of the innermost parts of the disk are intrinsically non-Keplerian.

\section{Discussion}

The ALMA observations towards IRAS 16547 revealed a rich variety of molecular species related 
to both continuum sources (see Table 1). This molecular emission is present at different scales 
and temperatures. The molecules presented in Figure 3 (SO$_2$ or OCS) trace, at scales about 
10$^4$ AU, a clear velocity gradient with a similar orientation to the thermal jet.  Some other molecules also detected in this observation 
 ({\it e.g. } CH$_3$OH, CH$_2$NH, H$_2$CCNH) with similar excitation temperatures in the upper states 
 (E$_u$ $\sim$ 300 K) are tracing similar structures at scales of some 10$^4$ AU.  
 On the other hand, the maximum velocity that an object with mass $m$ can achieve 
 under a gravitational field of an object with mass $M$ can be inferred from a balance between gravity and centrifugal 
 forces: 
 
 $$
 \frac{mv^2}{R} - \frac{GMm}{R^2} = 0,
 $$ 
 
 \noindent
 where $R$ is the distance to the object of mass $M$ creating the gravitational field, $G$ is the gravitational constant, and $v$ is the velocity. 
 This yields a dynamical mass given by 
 
 $$
 M = \frac{Rv^2}{G}.
 $$
 
 \noindent
 Writing this equation in practical units, we obtain: 
 $$
 M[M_\odot]=1.13 \times 10^{-3} v^2 [km s^{-1}] R[AU].  
 $$
 
 \noindent
 If we assume that the gradient observed in Figure 3 is produced by a rotating 
 structure centered in IRAS 16547, we estimated a dynamical mass of 60 M$_\odot$.
 This mass corresponds to an early O-type protostar  with a bolometric luminosity of $\sim$ 10$^6$ L$_{\odot}$ \citep{pan1973}, which exceeds
 by one order of magnitude the bolometric luminosity of IRAS 16547 ($\sim$10$^5$ L$_{\odot}$). We then suggest that this
 gradient might be produced by the thermal jet that entrains the molecular gas, and that has a similar orientation.  Moreover,
 the emission of the high velocity wings from the SO$_2$ is very extended in the direction of the jet. We therefore conclude that the
 SO$_2$ emission is most likely tracing an outflow.  

At much more smaller scales ($\sim$ 1000 AU), molecules with high excitation temperatures (E$_u$ $\gtrsim$ 500 K) trace 
a compact rotating structure perpendicular to the orientation of the thermal jet (see Figure 4 \& 5). This group of molecules 
includes: SO$_2$ v2$=$1, HNCO v$=$0, CH$_3$SH, CH$_3$OH, etc. If we do a similar estimation as above, we obtain a dynamical 
mass of $\sim$ 26 M$_\odot$.This value for the mass is uncorrected by the inclination angle.  
This corresponds to a bolometric luminosity of 10$^{4-5}$ L$_{\odot}$, a similar bolometric luminosity
to that of IRAS 16547. Given the dimensions, the orientation, and the dynamical mass estimated from the velocity gradient traced by these
molecules, we suggest that all these molecules trace an rotating disk surrounding IRAS 16547.  
Similar results are found for the dynamical mass of the  protostar, if we assume 
that the water masers reported by \citet[][]{fra2009} trace a very compact disk with a similar P.A., and 
at much more smaller scales ($\lesssim$ 300 AU).  An upper limit mass to the disk 
is $\sim$ 6 M$_\odot$, as estimated from the dust thermal emission. In conclusion, removing the mass of the disk, 
the protostar in the middle has a mass of $\sim$ 20 M$_\odot$.

It is interesting to note that even the same molecular species, but in different transitions ({\it e.g.} SO$_2$ and SO$_2$ v$_2$=1) 
trace distinct components of IRAS 16547. For example, the SO$_2$ v=0 28$_{2, 26}$ - 28$_{1, 27}$ is tracing the outflow (see Figure 3), while the SO$_2$ v$_2$=1 8$_{2, 6}$--7$_{1, 7}$ is tracing the hot disk. This is because of the different excitation temperatures of the molecule 
at different transitions (see Table 1). 


\section{ Conclusions}

We have carried out submillimeter line and continuum observations made with ALMA of the massive protostar IRAS 16547.
The main conclusions of this study are as follows:

\begin{itemize}

\item In the 0.85 mm continuum band, the observations revealed two compact sources (with a separation of $\sim$ 2$''$), 
one of them associated with IRAS 16547$-$4247, and the other one to the west. Both sources are well resolved angularly, 
revealing a clumpy structure. 

\item At scales larger than 10,000 AU, molecules ({\it e.g.}, SO$_2$ or OCS) 
mostly with low excitation temperatures in the upper states (E$_k$ $\lesssim$ 300 K) are present in both millimeter continuum sources, and 
show a southeast-northwest velocity gradient of 7 $\kms$ over 3$''$ (165 $\kms$ pc$^{-1}$). 
We suggest that this gradient probably is produced by the thermal (free-free) jet emerging from this object with a similar orientation at the base.

\item At much smaller scales (about 1000 AU),
molecules with high excitation temperatures (E$_k$ $\gtrsim$ 500 K) are tracing a rotating structure elongated perpendicular to the orientation 
of the thermal jet, which we interpret as a candidate rotating disk surrounding IRAS 16547$-$4247. From Keplerian arguments, we estimate a mass of about 20 M$_\odot$ for the central star.

\end{itemize}

\section*{Acknowledgements}

L.A.Z, A. P., R. G. and L. F. R. acknowledge the financial support from DGAPA, UNAM, and CONACyT, M\'exico. 
This paper makes use of ALMA data: ADS/JAO.ALMA\#2011.0.00419.S. ALMA is a partnership of ESO (representing its member states), 
NSF (USA), and NINS (Japan), together with NRC (Canada) and NSC and ASIAA (Taiwan), in cooperation with the 
Republic of Chile. The Joint ALMA Observatory is operated by ESO, AUI/NRAO, and NAOJ. 


\begin{thebibliography}{}
\bibitem[Araudo et al.(2007)]{ara2007} Araudo, A.~T., Romero, G.~E., Bosch-Ramon, V., \& Paredes, J.~M.\ 2007, \aap, 476, 1289
\bibitem[Cernicharo et al.(2012)]{cer2012} Cernicharo, J., Marcelino, N., Roueff, E., et al.\ 2012, \apjl, 759, L43
\bibitem[Franco-Hern{\'a}ndez et al.(2009)]{fra2009} Franco-Hern{\'a}ndez, R., Moran, J.~M., Rodr{\'{\i}}guez, L.~F., \& Garay, G.\ 2009, \apj, 701, 974 
\bibitem[Galv{\'a}n-Madrid et al.(2010)]{gal2010} Galv{\'a}n-Madrid, R., Zhang, Q., Keto, E., et al.\ 2010, \apj, 725, 17
\bibitem[Garay et al.(2003)]{gar2003} Garay, G., Brooks, K.~J., Mardones, D., \& Norris, R.~P.\ 2003, \apj, 587, 739 
\bibitem[Garay et al.(2007)]{gar2007} Garay, G., Mardones, D., Bronfman, L., et al.\ 2007, \aap, 463, 217
\bibitem[Gibb et al.(2000)]{gib2000} Gibb, E., Nummelin, A., Irvine, W.~M., Whittet, D.~C.~B., \& Bergman, P.\ 2000, \apj, 545, 309
\bibitem[Gooch(1996)]{goo96} Gooch, R.\ 1996, Astronomical Data Analysis Software and Systems V, 101, 80
\bibitem[Hern{\'a}ndez-Hern{\'a}ndez et al.(2014)]{her2014} Hern{\'a}ndez-Hern{\'a}ndez, V., Zapata, L., Kurtz, S., \& Garay, G.\ 2014, \apj, 786, 38
\bibitem[Hildebrand(1983)]{hil1983} Hildebrand, R.~H.\ 1983, Quarterly Journal of the Royal Astronomical Society, 24, 267 
\bibitem[Hoffman(2012)]{hof2012} Hoffman, I.~M.\ 2012, \apj, 759, 76
\bibitem[Jim{\'e}nez-Serra et al.(2012)]{jim2012} Jim{\'e}nez-Serra, I., Zhang, Q., Viti, S., Mart{\'{\i}}n-Pintado, J., \& de Wit, W.-J.\ 2012, \apj, 753, 34 
\bibitem[Klaassen et al.(2009)]{kla2009} Klaassen, P.~D., Wilson, C.~D., Keto, E.~R., \& Zhang, Q.\ 2009, \apj, 703, 1308
\bibitem[Kolesnikov{\'a} et al.(2014)]{kol2014} Kolesnikov{\'a}, L., Tercero, B., Cernicharo, J., et al.\ 2014, \apjl, 784, L7
\bibitem[Krumholz et al.(2009)]{kru2009} Krumholz, M.~R., Klein, R.~I., McKee, C.~F., Offner, S.~S.~R., \& Cunningham, A.~J.\ 2009, Science, 323, 754  
\bibitem[Kuiper et al.(2010)]{kui2010} Kuiper, R., Klahr, H., Beuther, H., \& Henning, T.\ 2010, \apj, 722, 1556 
\bibitem[Linke et al.(1979)]{lin1979} Linke, R.~A., Frerking, M.~A., \& Thaddeus, P.\ 1979, \apjl, 234, L139
\bibitem[McMullin et al.(2007)]{mc2007} McMullin, J.~P., Waters, B., Schiebel, D., Young, W., \& Golap, K.\ 2007, Astronomical Data Analysis Software and Systems XVI, 376, 127 
\bibitem[Moscadelli \& Goddi(2014)]{mos2014} Moscadelli, L., \& Goddi, C.\ 2014, \aap, 566, A150
\bibitem[M{\"u}ller et al.(2005)]{mul2005} M{\"u}ller, H.~S.~P., Schl{\"o}der, F., Stutzki, J., \& Winnewisser, G.\ 2005, Journal of Molecular Structure, 742, 215 
\bibitem[Naranjo-Romero et al.(2012)]{nar2012} Naranjo-Romero, R., Zapata, L.~A., V{\'a}zquez-Semadeni, E., et al.\ 2012, \apj, 757, 58
\bibitem[Ossenkopf \& Henning(1994)]{oss1994} Ossenkopf, V., \& Henning, T.\ 1994, A\&A, 291, 943
\bibitem[Palau et al.(2011)]{pal2011} Palau, A., Fuente, A., Girart, J.~M., et al.\ 2011, \apjl, 743, L32 
\bibitem[Palau et al.(2013)]{pau2013} Palau, A., S{\'a}nchez Contreras, C., Sahai, R., S{\'a}nchez-Monge, {\'A}., \& Rizzo, J.~R.\ 2013, \mnras, 428, 1537
\bibitem[Panagia(1973)]{pan1973} Panagia, N.\ 1973, \aj, 78, 929
\bibitem[Pickett et al.(1998)]{pic1998} Pickett, H.~M., Poynter, R.~L., Cohen, E.~A., et al.\ 1998, JQSRT, 60, 883
\bibitem[Peters et al.(2010)]{pet2010} Peters, T., Banerjee, R., Klessen, R.~S., et al.\ 2010, \apj, 711, 1017
\bibitem[Qiu et al.(2012)]{qiu2012} Qiu, K., Zhang, Q., Beuther, H., \& Fallscheer, C.\ 2012, \apj, 756, 170
\bibitem[Ricci et al.(2010)]{ric2010} Ricci, L., Testi, L., Natta, A., et al.\ 2010, \aap, 512, A15
\bibitem[Rodr{\'{\i}}guez et al.(2008)]{rod2008} Rodr{\'{\i}}guez, L.~F., Moran, J.~M., Franco-Hern{\'a}ndez, R., et al.\ 2008, \aj, 135, 2370  
\bibitem[Wang et al.(2012)]{wan2012} Wang, K.-S., van der Tak, F.~F.~S., \& Hogerheijde, M.~R.\ 2012, \aap, 543, A22
\bibitem[Zapata et al.(2009)]{zap2009} Zapata, L.~A., Ho, P.~T.~P., Schilke, P., et al.\ 2009, \apj, 698, 1422
\bibitem[Zapata et al.(2010)]{zap2010} Zapata, L.~A., Tang, Y.-W., \& Leurini, S.\ 2010, \apj, 725, 1091 
\bibitem[Zapata et al.(2011)]{zap2011} Zapata, L.~A., Rodr{\'{\i}}guez-Garza, C., Rodr{\'{\i}}guez, L.~F., Girart, J.~M., 
\& Chen, H.-R.\ 2011, \apjl, 740, L19 
\bibitem[Zinnecker \& Yorke(2007)]{zin2007} Zinnecker, H., \& Yorke, H.~W.\ 2007, \araa, 45, 481
\end{thebibliography}
\end{document}